\documentclass[letterpaper]{article}
\usepackage{aaai}
\usepackage{times}
\usepackage{helvet}
\usepackage{courier}
\usepackage{graphicx}
\usepackage{subfig}
\usepackage{url}
\title{Like, Comment, Repin: User Interaction on Pinterest}
\author{Bluma Gelley  \\
New York University \\
Polytechnic School of Engineering \\
Brooklyn, NY \\
bgelley@nyu.edu \\
\And 
Ajita John \\
Avaya Labs\\
Basking Ridge, NJ, USA\\
ajita@avaya.com}

\begin{document}

\maketitle

\begin{abstract}
We present the results of a study of the Pinterest activity graph. Pinterest is a relatively new and extremely popular content-based social network. Building on a body of work showing that the hidden network whose edges are actual interactions between users is more informative about social relationships than the follower-following network, we study the \textbf{\textit{activity graph}} composed of links formed by liking, commenting, and repinning, and show that it is very different from the follow network. We collect data about 14 million pins, 7 million repins, 1.6 million likes, and several hundred thousand users and report interesting results about social activity on Pinterest. In particular, we discover that only 12.3\% of a user's followers interact with their pins, and over 70\% of activity on each user's boards is done by non-followers, but on average, followers who are active perform twice as many actions as non-followers. 

\end{abstract}

\section{Introduction}
\label{sec:introduction}
	Pinterest is an Online Social Network (OSN) centered around the curation and sharing of visual content. Since its inception in 2010, it has grown extremely rapidly, reaching 10 million monthly unique visitors faster than any OSN ever\footnote{Comscore report, March 2013}, and boasting 70 million users by July 2013. A 2013 Pew Internet Survey \cite{pew} found that 21 \% of all Internet users in the US use Pinterest. The site focuses on the curation and sharing of highly visual content by its users, using the metaphor of virtual `pinboards' on which images and other media are `pinned'. Users can follow other users, or pinners, to see all of their content in their own home feed. Pinterest has attracted much attention from marketers, due to its ``aspirational'' nature, with users using the site to find and share products and services that they would like to buy.  By late 2013, Pinterest was driving 20\% of all social network referrals to purchasing sites, second only to Facebook\footnote{Comscore, October 2013}. Pinterest's popularity and the marketing opportunities it presents makes it an interesting and important subject of study. 
	On most online social networks, followers are widely perceived as extremely important. Follower (or friend, on networks with undirected edges) count is often seen as synonomous with the size of a user's audience and even the amount of influence they have; users with large numbers of followers are viewed as very influential and are often sought after by marketers to promote their products \cite{huberman2008social}.
	Standard OSN analysis uses the network formed by linking users with their followers to derive information about relationships between users. A large body of work, however, has shown that this network, which we refer to as the \textbf{\textit{follow graph}}, contains incomplete information about users and their relationships, since it does not capture active interaction. Studies of Facebook \cite{wilson2009user}, Twitter \cite{huberman2008social}, and Cyworld \cite{chun2008comparison} all found that the hidden network formed by interactions between users is very different than the follow graph. In particular, users tend to interact with only a small subset of their followers \cite{marlow2009maintained,huberman2008social}; thus, relationship strength, and by implication, influence, are difficult to predict using the follow graph alone \cite{cha2010measuring,suh2010want,romero2011influence}. To augment the static follow graph, the above works use the \textbf{\textit{activity graph}}, a graph representation of the users in the social network where the directed edges between users are made up of active interactions between them, rather than passive follow links. Two users A and B who are connected in the follow graph (e.g., A follows B) will only be connected in the activity graph if there was an interaction of some sort between them. This relationship is called an \textbf{\textit{activity link,}} to distinguish it from the \textbf{\textit{follow links}} that form the edges in the follow graph. For some OSN's, such as Facebook, where users can only interact with their friends, the activity graph is a proper subset of the follow graph. On Pinterest, users can interact with all other users on the site, so the Pinterest activity graph can potentially be completely different than the follow graph. We define an activity link on Pinterest as a like, comment, or repins. These actions are described in section \ref{sec:pintdesc}.

	It is clear, then, that in many OSN's, the implicit activity network provides additional, and perhaps more accurate, information about relationships between users than the static follow graph. Is this true of Pinterest as well? Unlike social networks such as Facebook and Cyworld, Pinterest is centered around content, with all activity revolving around the content. Even microblogging services like Twitter, which also involve content, are much more social than Pinterest. Users tend to tweet about themselves, and to follow other tweeters they are interested in hearing from. On Pinterest, the focus is on content rather than the content creators; pinners are encouraged to follow boards (see section \ref{sec:pintdesc})whose topics they are interested in, rather than users. This difference can easily be seen in the difference between the highest-ranked users on Twitter and on Pinterest. Nearly all of the top 100 users with the highest follower counts on Twitter are celebrities or news sites \footnote{http://www.twitaholic.com}. On Pinterest, the top 10 are all ``ordinary'' users whose content, not fame, is what won them followers. The same is true of Instagram; while it is also a visual image-sharing site, its structure promotes building relationships with other users through their images - in many ways, a visual-based Twitter.

	Given the primacy of content on Pinterest and its function as a motivation for following, it may be reasonable to assume that the activity network on Pinterest closely parallels the follow network. We find, however, that this is not the case. In this paper, we sample the implicit activity network on Pinterest and analyze various aspects of user interaction on the site. We find that the distributions of activity across pins, boards, and users all follow a power law, and that repins are about 4 times more common than likes and 150 times more common than comments. We report several differences between the follow and activity graph. In particular, we find that on average, over 70\% of all activity on a user's boards is not done by their followers. Conversely, only 12.3\% of an average user's followers interact with the user's pins. Those followers who are active, however, perform twice as many actions, on average, as non-followers. Though a user's number of followers is strongly correlated with their number of pins, the rate of activity on their boards is not; for most users, activity and number of followers are very weakly correlated. To the best of our knowledge, this is the first study of the Pinterest activity network.  

	The remainder of this paper is structured as follows. First, we provide an overview of Pinterest in section 2. In section 3, we discuss related work. Our sampling and data collection methods are described in sections 4 and 5, along with our dataset. In section 6, we report the results we obtained from our analysis of the dataset, and discuss their implications in section 7. We conclude in section 8, and propose some ideas for future work.

\section{Overview of Pinterest}
\label{sec:pintdesc}
	Pinterest describes itself as "A tool for collecting and organizing things you love." It's billed as a "virtual pinboard" service, where users can easily 'pin' digital content they find interesting or useful and share it with others. The central entity on Pinterest is the \textbf{pin}. A pin is an image or video, often accompanied by a caption. Pins can be uploaded by the user or reposted from somewhere else on the Web; these link back to the original source when clicked on. Users, or \textbf{pinners}, \textbf{pin} content onto their \textbf{boards} - pages, usually organized around a specific theme, where pins are laid out in an informal style reminiscent of a physical pinboard. Pinterest's trademark layout is designed for maximum visual appeal: pins are displayed in neat rectangles of varying heights in a grid pattern that continuously loads new content as the viewer scrolls down. Clicking on a pin opens it on a separate page with more detail.  

	User profiles on Pinterest are pretty basic: the only information required is a first name. Users can add a profile picture, a brief description, and their location, as well as links to Facebook, Twitter, and/or a personal website. Also displayed are the user's number of boards, pins, and likes, and their number of followers and following (other users the user follows). The rest of the profile page is devoted to the user's boards. Like pins, these are laid out as small rectangles in a grid, and  display several images from the board. 
	Every board on Pinterest belongs to a category. A continuously changing sample of pins from boards in each category are reposted on the category pages, accessible from a drop-down menu on the site header. There is also a "popular" page, where popular pins from around the site are displayed, and a search box for finding pins, boards, or users matching a search term. 
	As with any social network, Pinterest users can connect with other users by \textbf{following} them. Follow edges in Pinterest are directed; pinners can follow other users who do not follow them back. In addition, following someone does not require their permission. Since pinners often have boards on many different topics, users often follow only those boards which interest them \cite{zarro2013wedding}; board followers are counted in the user's general follower count on his/her profile page. New pins from followed boards or users show up in a follower's \textbf{home feed}, which they see when they open Pinterest. Users can also create and join group boards, some of which have thousands of pinners posting content to them. Each user is also allowed up to 3 secret boards, which are only visible to the owner(s).
	There are three types of active social interaction on Pinterest: likes, comments, and repins. According to the Pinterest help page, 'Like a pin when you want to say “Hey you! Neat idea!”' Liked pins are also saved to the "Likes" page in the user's profile, so pinners will often like pins that they want to be easily able to find later. Users can also comment on a pin; comments are displayed beneath the pin on the board. Finally, users can repin a pin to one of their own boards. Repinning is the equivalent of retweeting on Twitter or reblogging on Tumblr and is the mechanism by which pins can "go viral". Hovering over a pin on a board page displays the like and repin buttons, but to comment, a user must click on the pin to open it in a separate page. Pinterest encourages business to use its site by providing specialized business accounts with options that help businesses better market their products on Pinterest.
  
\section{Related Work}
\label{sec:relatedwork}
	Pinterest is a fairly new site, and its lack of an API has created an additional barrier to its study. A few analyses of Pinterest, however, have been published very recently. \cite{statoverviewpint} attempt to determine what drives user behavior on Pinterest by calculating the contribution that various factors (such as the gender and nationality of the original pinner) have to the likelihood of a pin's being repinned and the number of followers a user attracts. They also compare the language used by the same users on Pinterest and on Twitter and determine that there are significant differences. \cite{studyuserbehavorspint} study user behavior on Pinterest, but they confine their analysis to the static follow graph; they also study the content and categories of pins, particularly popular ones. In two somewhat overlapping papers, \cite{mittal2013pin} Mittal, et. al. analyze various aspects of Pinterest, including some user characteristics, the distribution of user locations, and pin sources. They also address privacy and copyright issues and find many instances of personal data leakage and copyright violations on Pinterest. Finally, they find that they can predict gender of Pinterest users with high accuracy. Gender differences in Pinterest are a popular topic of study; \cite{ottoni2013ladies}quantify the differences in Pinterest behavior between male and female users. \cite{chang2014specialization} also study gender, specifically, the types of content favored by and degree of specialization of the two genders. They also report that homophily - here defined as similarity in interests - has a large influence on repinning, but a smaller one on following.  \cite{boardrecs} build a model to automatically recommend boards that users might like. \cite{zarro2013wedding} studied Pinterest from a quantitative perspective through user interviews. Their findings confirm that users see Pinterest as a content provider rather than a social network.

	The concept of the implicit activity network in an online social network and the fact that it differs from the explicit follow network was first proposed by \cite{chun2008comparison}, who analyzed the topographical characteristics of both the follow and activity networks of Cyworld, a Korean OSN. They found that the one-way interaction network had a similar topology to the follow network, but the reciprocal "friends" network was quite different, more similar to known topologies of offline social networks than to the usual characteristics of OSNs. \cite{ahn2007analysis} had previously made a similar observation about the testimonial network on Cyworld, but did not extend their results to the concept of the activity graph in general.  \cite{wilson2009user} performed a very similar analysis on Facebook, referring to the implicit network as the \textit{interaction graph}.They found significant differences between the static follow graph and the interaction graph, once again finding that the interaction graph displays the small-world properties typical of OSN graphs to a lesser extent than the follow graph does. The Twitter interaction graph was studied by \cite{huberman2008social}, who compared "friends" (other users the user directed at least two tweets at) with declared followers and found that most users have many more followers than friends; that is, they interact closely with only a small subset of their followers. This disparity was confirmed in the case of Facebook by the Facebook Data Science team, who, with access to all of Facebook's user data, showed that the number of active reciprocal relationships per user was much smaller than the user's friend count \cite{marlow2009maintained}. In a similar vein, \cite{cha2010measuring} challenged common assumptions about influence by showing that follower count was not strongly correlated with influence. \cite{romero2011influence} performed a similar analysis, creating a method for calculating influence in Twitter that accounts for the high amount of passivity among users, which makes the identification of active content forwarders essential for the propagation of information; they state specifically that number of followers is not a good measure of influence. 

\section{Sampling}
\label{sec:sampling}
	As discussed in section \ref{sec:introduction}, we concentrated on the \textit{Activity Graph} rather than the static follower graph. Instead of crawling follower edges, we followed activity links between users (though we did collect follow links as well for crawled users). As is common in OSN analysis, analyzing the entire Pinterest graph was impractical; we therefore did our analyses on a sample of the network. Since unique ids on Pinterest are text-based, random sampling was difficult; it was also undesirable in our case, since we wanted to collect as well-connected a graph as possible. Instead, we used modified Breadth-First Search (BFS) (that is, Snowball Sampling \cite{goodman1961snowball}) on the full graph $G(V,E) $ to collect the sample $S(V',E'),V' \subset V, E' \subset E $ beginning from several randomly chosen seeds and moving outward by selecting a random subset of edge clusters and crawling all edges in each cluster. This is accomplished by randomly choosing k boards of each crawled user, and then crawling all activity links on each board.  
\label{bfsjustification}
 	While BFS, and by extension snowball sampling, \cite{illenberger2008approach} have been shown to be biased towards high-degree nodes \cite{kurant2010bias,gjoka2010walking,lee2006statistical}, we chose BFS because we wished to capture the interactions between tightly connected groups of users. This goal is well served by BFS, which excels at fully covering small regions of a graph \cite{kurant2011towards}. BFS has been used for many analyses of OSNs, among them \cite{ahn2007analysis,mislove2007measurement}, and \cite{wilson2009user}. Although our method of crawling only k boards for each user $ v_{i} $ yields an incomplete view of the neighborhood of each node in S, technical (see section \ref{crawlingchallenges}) and time limitations forced a tradeoff between number of users crawled and completeness of the data gathered for each user. We chose k =5 as the value that best optimizes this tradeoff.  Likewise, we limited the number of pins collected per board to the first 300, since collecting more than that was prohibitively expensive, and 90\% of boards have 300 or fewer pins. On a similar OSN graph, \cite{bonneau2009eight} were able to estimate many graph properties using, as we do, only a random sample of k edges from each node, even with very small values of k. Due to the expense of crawling large follower lists, we limited our analysis to users with 10,000 followers or fewer.
 
\section{Data Collection}
\label{crawlingchallenges}
	Collecting data from Pinterest was an extremely complex challenge. Unlike most large social network services, Pinterest does not have an API for downloading user data. We therefore created a crawler to download all publicly available data visible on the site itself. As of this writing, Pinterest does not have any privacy controls, so all users and their boards (with the exception of secret boards) are publicly visible. Pinterest's trademark design, which utilizes infinite scrolling (a design technique where new content is continuously loaded as the user scrolls down), and its method of storing a single user's data on many separate pages, makes automatic data collection difficult. Crawling a single user with just an average number of boards, pins, followers, and social interaction can require over 1700 server calls!  To add to the difficulty, we found that the actual number of an entity's followers, pins, or likes/comments/repins frequently differed, often significantly, from the number displayed on the corresponding page. This required us to use other validation methods to insure that the data collected was accurate and complete. The lack of an API also meant that some data was not available to us; for instance, we had no access to any temporal data. Shortly before we began this project, Pinterest introduced a complete redesign of the site, one part of which makes standard http-request-based crawling impossible. Instead, we used Selenium Webdriver, a browser automator that can simulate the action of scrolling a page. Browser automation is significantly slower than issuing and parsing http requests, so the change was effectively a strict rate-limiting mechanism. The redesign included many changes aimed at altering the way that users interact with the site. Our dataset, therefore, contains important information about social interaction in the `new Pinterest,' and also includes new features such as Rich Pins and Place Boards.

\subsection{Crawler Architecture}
\begin{figure}[tbph]
\centering
\includegraphics[width=0.9\linewidth, height=0.35\textheight]{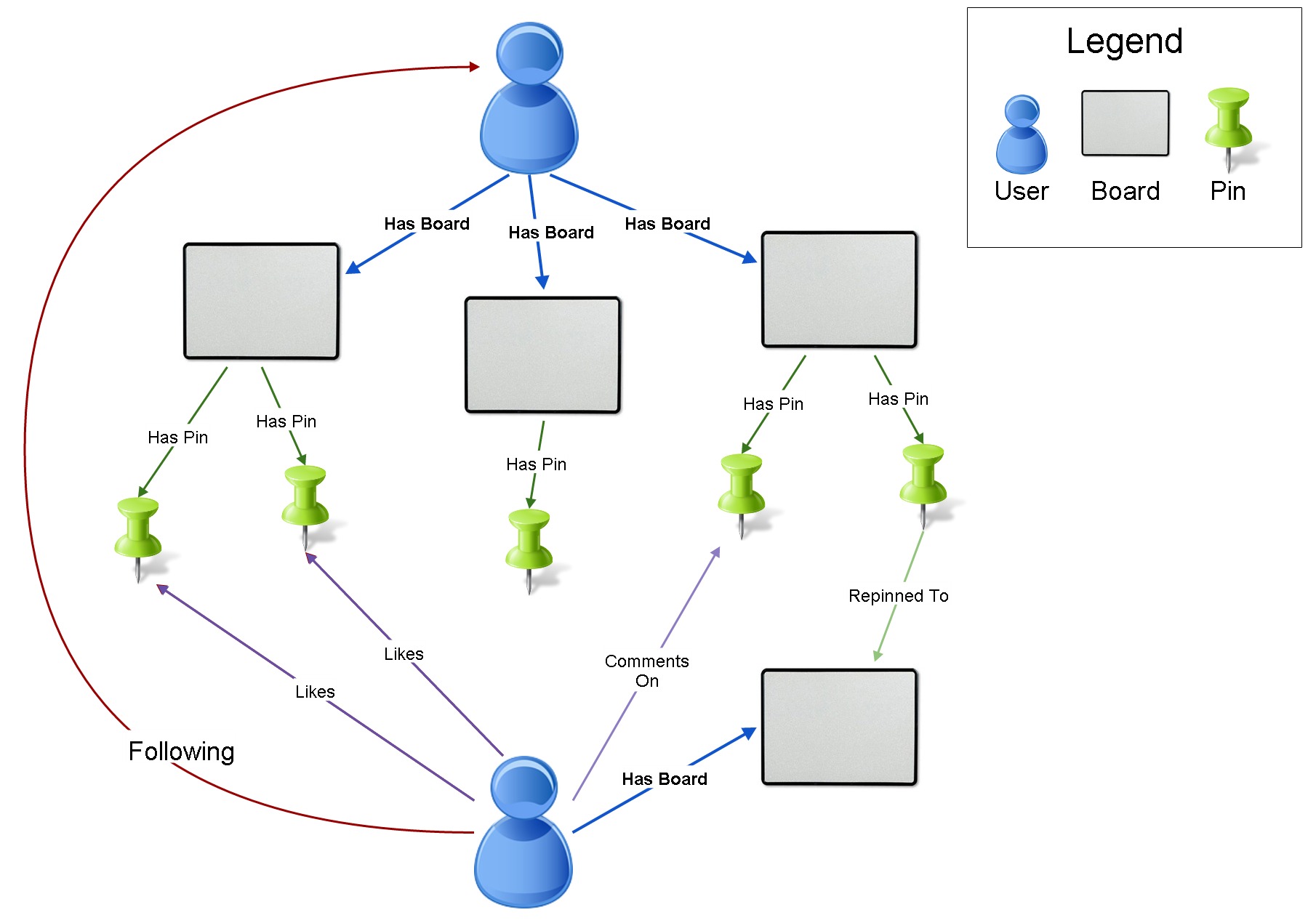}
\caption{The Pinterest data model}
\label{fig:datamodel1}
\end{figure}

We utilize a multithreaded crawler architecture for data collection. The controller maintains a shared FIFO queue of usernames to crawl as well as a set containing all usernames already crawled. Each thread contains a separate browser instance, which handles the task of visiting the profile page and 5 board pages for each user and downloading all available data. The next step is accessing the separate pages (see section \ref{crawlingchallenges}) storing any likes and repins for each pin. Like pages display partial profile information for all users who liked the pin, and repin pages contain information about the board the pin was repinned to. User, board, pin, and like/repin/comment data is all added to the database as entities connected by relationships, as shown in the data model in Figure \ref{fig:datamodel1}. Finally, we download all followers for each crawled user in a separate process and add them to the graph database as well. We ran the crawler for 5 weeks in December 2013 and January 2014 and collected the data described in Figure \ref{fig:data}.

\subsection{Data}

 We collect all available data about each of these entities and their relationships. Since our data forms a network structure, it is well suited for storage in a graph database, which represents entities and the relationships between them as nodes and edges in a graph. We use the Neo4j database system for our storage, and its Cypher query language for data extraction from the graph. \cite{neo4j} The details of our dataset are shown in Table \ref{fig:data}. \begin{table} 
    \begin{tabular}{|l|l|}
    \hline
    Dataset Details                      & ~              \\ \hline
    Crawled Users (with boards and pins) & 31,359        \\ \hline
    Users (partial data)                 & 4.5 million    \\ \hline
    Total Users touched                  & 5.4 million      \\ \hline
    Crawled Boards                       & 150,000         \\ \hline
    Boards (partial data)                & 200,000       \\ \hline
    Total Boards Touched                 & 5.1 million      \\ \hline
    Total Pins Crawled                   & 14 million    \\ \hline
    Total Repins                         & 7 million       \\ \hline
    Total Likes                          & 1.56 million              \\ \hline
    Total Comments                       & 47,557            \\ \hline
    \end{tabular}
    \caption{Data Description}
    \label{fig:data}
\end{table}

\section{Analysis}
\subsection{Data Distributions}
\begin{figure}[tbph]
\centering
\includegraphics[width=0.7\linewidth]{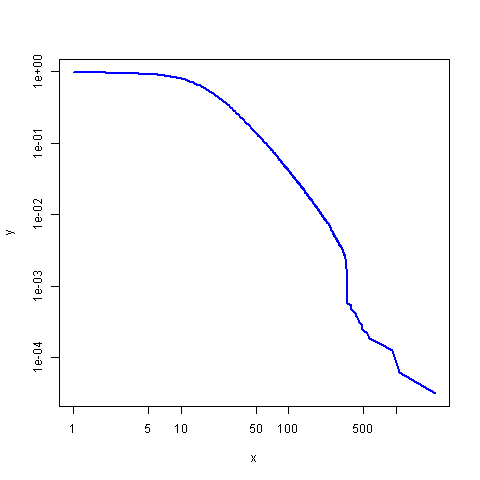}
\caption{The cumulative frequency distribution of the number of boards per user (log scale).}
\label{fig:dist1}
\end{figure}

\begin{figure}[tbph]
\centering
\includegraphics[width=0.7\linewidth]{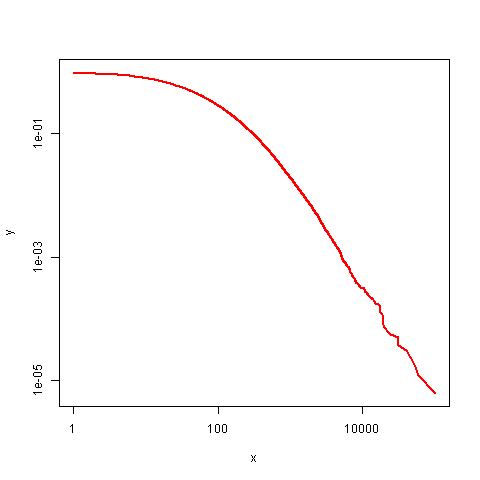}
\caption{The cumulative frequency distribution of the number of pins per board (log scale).}
\label{fig:dist2}
\end{figure}

	We first show the distributions of two types of content on Pinterest. Figures \ref{fig:dist1} and \ref{fig:dist2} show the distributions of the number of boards belonging to each user and the number of pins on each board. Both distributions follow a power law, as has been demonstrated for many other distributions related to OSNs and the web in general \cite{mislove2007measurement}. 


\subsection{Activity Links}
	As discussed in the introduction, activity links have been shown to be a better indicator of interaction than static following links. On Pinterest, activity links take the form of repins, likes, and comments, (see \ref{sec:pintdesc} for details), while follow links connect users and their followers (or followers of their boards). \begin{figure}[tbph]
\centering
\includegraphics[width=0.9\linewidth,height=0.4\textheight]{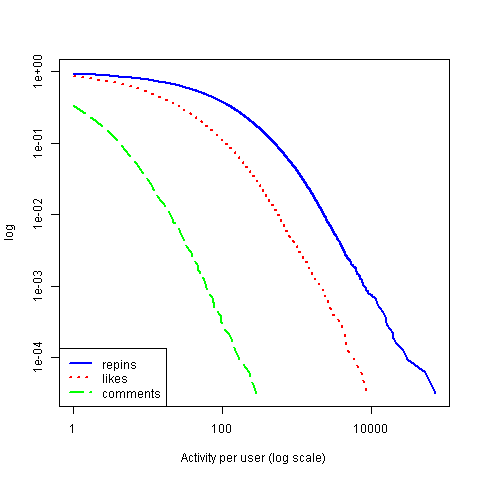}
\caption{Cumulative Distribution of Activity per User}
\label{fig:activityperuser}
\end{figure}

	Figure \ref{fig:activityperuser} shows the distribution of likes, comments, and repins across users, for the 5 boards/user that we crawled. Two things are immediately evident: the first is that repins are by far the most common type of activity, making up 81\% of total activity and, on average, 79\% of activity on each board. Likes are the next most frequent, and comments the least. Comments, in fact, are extremely rare; just .3\% of pins have even one comment. By contrast, 15.2\% of the pins in our dataset were repinned at least once. 
The second is the shape of the distribution. The three distributions fit power laws, as is clear from the plot. Clearly, some users are very successful at inspiring activity on their pins, while others have next to no activity at all. This would seem to be a function of the difference in number of followers between different users, but we will see in section \ref{follsandsocact} that this is not the case. The distribution of activity per pin is even more skewed; there are a few pins that tend to be interacted with repeatedly while others are ignored.

\subsection{Pins, Followers, and Activity Links}
	We next examine the relationship between a user's number of pins and the rate of activity on their pins. The correlation is a very weak one, with a Spearman's $\rho$ of 0.32. We compare to the follow graph by correlating the number of a user's pins with their number of followers. This correlation is a strong one, with $\rho =  0.78$. It may be that people with more followers are more motivated to post, or conversely, that users with more pins garner more followers. The lack of correlation with amount of activity, however, is very interesting. This would seem to contradict \cite{wu2009feedback}'s finding of a positive feedback loop between feedback on posts and posting frequency. Do pinners see only follower count as a positive feedback measure and ignore activity? These questions require much more study to answer, but they are extremely intriguing. Similar results have actually been reported for Twitter: \cite{suh2010want} found a linear relationship between retweets and followers, but little correlation between numbers of tweets and retweets. For a direct comparison, we correlated just the repin rate with the number of pins. Our results are similar to what Suh, et. al. report for retweets: pins and repins were barely correlated, ($\rho = 0.34 $), while the number of pins and followers have a strong relationship, as reported above. 

\subsection{Followers and Activity - Correlation}
\label{follsandsocact}
\begin{figure}[tbph]
\centering
\includegraphics[width=0.7\linewidth]{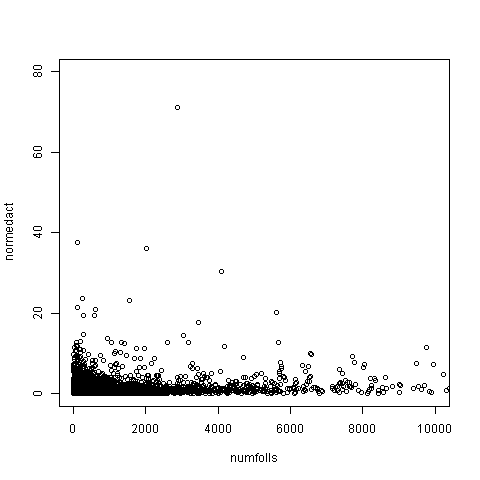}
\caption{Followers and Activity for All Users}
\label{fig:corrfollsandact}
\end{figure}

\begin{figure}[tbph]
\centering
\includegraphics[width=0.7\linewidth]{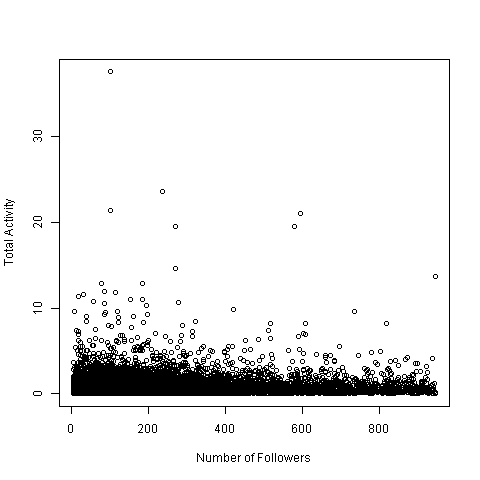}
\caption{Followers and Activity for the middle 85\% of users by follower count}
\label{fig:corrfollsandactmiddle}
\end{figure}

	The difference between the static follow graph and the activity graph can be seen quite starkly in the relationship between a user's number of followers (follow graph) and the rate of activity on their pins (activity graph). (Figure \ref{fig:corrfollsandact}) Conventional wisdom would expect there to be a strong correlation; the more people see a pin, the higher the likelihood that some will interact with it. We find, however, that this is not the case. The correlation between the number of followers and the activity rate is moderate ($\rho = .55$). On closer investigation, however, we find that this correlation is heavily influenced by outliers; when we calculate the follower-activity correlation for only those users between the 10th and 95th percentiles by follower count, the correlation for these ``average users'' is much weaker - just $\rho = .44$. (Figure \ref{fig:corrfollsandactmiddle})There are two possible explanations for this finding. One is that not all of a user's followers see the user's content. The other is that only a small percentage of those who see a pin will act on it. Both of these options are included in \cite{romero2011influence}'s definition of passivity; our finding corroborates theirs that the number of followers is a poor predictor of the rate of reaction to a post.   

	Interestingly, Pearson's r is much lower than $\rho$ in both the above cases (r = .23 for the average users). This suggests that even the weak correlation that does exist is monotonic rather than linear. That is, the amount of incoming activity does not increase linearly with the number of followers, but rather has just a very general upward trend that may only be noticeable with very large numbers of followers.
	There is also an interesting 'bump' at the y axis, consisting of a surprisingly large number of users who have no followers at all but nevertheless have activity on their pins. These users make up 1.4\% of our dataset and strengthen the case for the importance of the activity graph. These are users who would have been dead ends in the follow graph, but they have a reasonable (and in some cases, large) amount of influence on other users, as is clear from the amount of interaction their pins received. Conversely, there are significant numbers of users all along the x axis who have little to no activity but do have followers - in many cases, large numbers of them. Approximately 16\% of users have followers but no activity. 4\% of those have over 100 followers. These users may have been considered heavily connected when only the follow graph is taken into account, but the activity graph shows that they are not very successful at inspiring interaction and reposting of their content. (Note that since we only have a sample of boards, it is possible that these users have some amount of activity on their non-sampled boards. However, given our random sampling methods, we can be fairly confident that there is a non-trivial number of users with followers and no activity, even if it is not as high as the number we report.)  

\subsection{Active Followers}
\begin{figure}[tbph]
\centering
\includegraphics[width=0.7\linewidth]{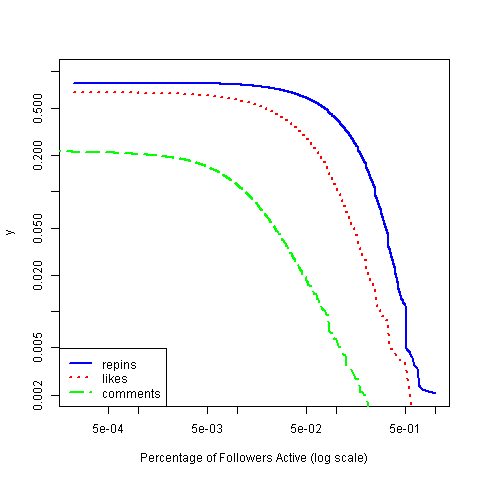}
\caption{Percentage of Followers who Interacted, by type of activity (log-log cumulative frequency)}
\label{fig:densitypercentfollsinteract}
\end{figure}

	One of the main arguments for the superiority of activity graphs over follow graphs as a measure of actual relationships is that only a small percentage of users' followers actively interact with them. This has been shown for Facebook, Twitter, and Cyworld (see section \ref{sec:relatedwork} We find that this is true of Pinterest as well: on average, only 12.3\% of a user's followers have ever engaged in a single interaction (like, comment, or repin), with any of the user's pins. The distribution for all users with any activity on their crawled boards, separated by type of activity, is shown in Figure \ref{fig:densitypercentfollsinteract}. Likes and comments once again have a more uneven distribution than repins; so not only do those users who interact comment and like much less than they repin, there are also fewer users who engage in either of the two activities than who repin. Even repins, however, are only done by a few followers: very few users have repins from more than 20\% of their followers.  

\subsection{Degree}
\begin{figure}[tbph]
\centering
\includegraphics[width=0.7\linewidth]{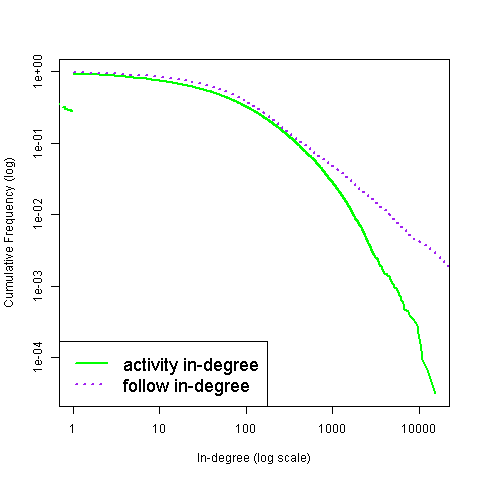}
\caption{In-degree Distribution for Follow and Activity Graphs. The in-degree for the follow graph is the number of followers; for the activity graph, it's the number of unique users interacting with a user.}
\label{fig:indegree}
\end{figure}

Node degree is a common measure used for evaluating graphs. It is also particularly important to us since it quantifies the difference between active and passive social links. We calculate the in-degree for both the follow graph and the activity graph and compare the results. For the follow graph, in-degree is the number of followers. In the activity graph, in-degree is the number of unique users who interacted with a user. (Note that activity graph in-degree is a separate measure from the percentage of a user's followers who interacted discussed in the last section, since in-degree includes all interacting users, not just followers.)The two distributions are shown in \ref{fig:indegree}; both are clearly power-law.
 
\subsection{Activity by Followers and Non-followers}
\begin{figure}[tbph]
 \centering
 \includegraphics[width=0.7\linewidth]{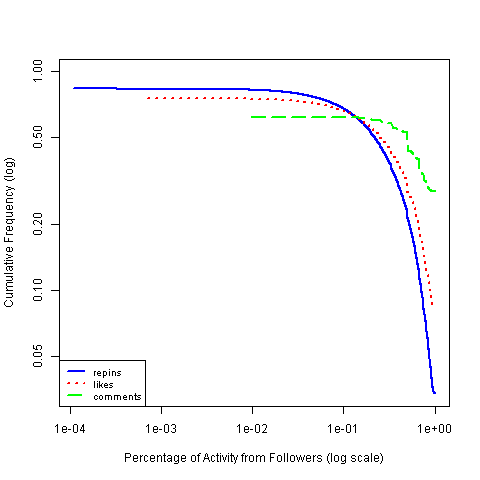}
 \caption{Percent of Activity on Each User's Boards Done By The User's Followers (Cumulative Distribution - log scale)}
 \label{fig:percentactfollsdensity}
 \end{figure}

	Unlike Facebook and similar OSN's, Pinterest is an open network - that is, users are not limited to interacting with their friends. Any user can like, comment on, or repin any other user's pins, without having to follow that user. In this sense, Pinterest is similar to Twitter, where anyone can see, respond to, and retweet anyone else's tweets. Pinterest is even more open than Twitter, since Twitter allows users to make their tweets visible to only their friends; this allows for the creation of small, closed circles of friends who use Twitter to foster their relationship rather than solely to share content. On Pinterest, all boards are publicly visible; secret boards are limited to just three and cannot even be viewed by followers. In this manner, Pinterest emphasizes its goals of content curation and sharing over social interaction and creating relationships.
	This openness provides a unique opportunity to help answer some important questions about Pinterest. How do users find content that they like? Do they stick primarily to their home feeds, and therefore only see (and potentially interact with) content from the users they follow? Or do they explore other boards as well, perhaps through the "Popular" or "Everything" feeds, the category pages, or the search box? (See section \ref{sec:pintdesc}.) Our data shows that users seem to be getting a large amount of the content they view from sources other than the users they follow. For the majority of users, only a small percentage of the interaction on their boards is from their followers - the median is just 24\%. The remaining 76\% or so are "drive-bys" - likes, comments, and repins done by users who do not want to be shown all of the user's content (or even one of their boards), but are interested in individual pins enough to interact with them. The numbers are not trivial: the median number of unique non-followers who interact with a user's board is 34. On average, non-following interacters make up a full 78\% of any user's unique interacters, but the average number of interactions done by each active non-follower is just 1.4. In contrast, the average active follower generates 3.4 interactions, more than twice as much. Figure \ref{fig:percentactfollsdensity} shows the distribution of percentage of repins, likes, and comments done by followers, for each user with at least one instance of the activity on their crawled boards. 
\section{Discussion}  
	Our data paints an intriguing picture of user interaction on Pinterest. While overall, there are 58 interactions (likes, comments, or repins) for every 100 pins, only 17.7\% of pins have even a single interaction, thanks to the skew in the distribution of activity per pin. Repins are by far the most common, about 4 times more common than likes and nearly 150 times more than comments. We attribute this to the nature of Pinterest itself, where the primary goal is discovering and curating content (i.e. pins), with social interaction coming in a distant second. Comments by nature involve far more social interaction than repins; repinning just means that the repinner wants the content for herself, while commenting is direct communication with the original pinner.
	The distributions of both followers and social activity per user follow power laws, suggesting that some users are much more effective than others at both collecting followers and fostering interaction on their boards. The amount of activity on a user's board, for those users in the "middle class" by follower count, however, is only weakly correlated with their number of followers and definitely does not increase linearly as more followers are added. The correlation is somewhat stronger when taking all users into account. It seems, then, that having huge numbers of followers does tend to result in a large amount of interaction, but going from a moderate number of followers to a slightly larger moderate number does not have very much of an impact. This is an important finding given the large amount of effort people put into attempts to increase their follower count. 
	Surprisingly, most of the activity on an average user's boards is not done by their followers. The average follower, however, contributes twice as much as the average non-follower. Conversely, only 12\% of a user's followers interact with any of the user's pins. This suggests a model of interaction where a user follows a number of other pinners (or one or more of their boards) and passively consumes their pinned content via her own home feed, while actively interacting with only a small subset of them. Users also find and view a significant amount of content from outside their home feed, perhaps through the "Popular" and "Everything" pages or the category pages, all of which contain a sample of pins from their respective topics, or by searching. It is also possible that users navigate Pinterest by following links from pins in their home feed to find content from pinners they do not follow. This can be done by following links to boards where pins they view or pin have been repinned to, or repinned from. Presumably, some of these newly discovered boards are interesting enough to motivate the user to follow them. Others, however, seem to have only a few pins that the user enjoys. The user takes these back to her own board by repinning, or has a brief interaction by liking or commenting, and then leaves. Obviously, this model requires more study and verification in order to be useful, but it is an interesting model strongly supported by our data.

\section{Conclusion and Future Work}
 	In this paper, we study social interaction on Pinterest using its activity graph, the network formed by activity links (repins, likes, and comments) between users, and compare it to the static follow graph. We find that the distributions of nearly everything on Pinterest - boards per user, pins per board, activity per pin, and many others as well, follow power laws, similar to those found repeatedly for other OSNs. We find that a user's incoming interaction rate is not well-correlated with their number of followers, nor with their number of pins. We also discover that the majority of repins, likes, and comments on a user's pins are not done by their followers, and that only a small percentage of a user's followers interact with the user's content. These results show that there is a large amount of information about user interactions available in the activity graph that is not visible in the follow graph. Along with our other findings, they provide strong support for carefully examining assumptions about number of followers being an accurate measure of influence on a social curation site. Given some of our findings, these assumptions may not be correct.
 	In this paper, we have mostly discussed characteristics of user interaction that can be learned from the activity graph. In the future, we would like to collect a larger, more fully connected dataset and examine various graph properties of the activity graph itself. We hope that this analysis will shed more light on Pinterest and the unique dynamics of user activity on a content-centered social network site. In particular, we are interested in the study of influence on Pinterest. Influence in social networks is an important topic of study because of the opportunities it presents for those who want to maximize the spread of ideas or advertising. Given the findings we present in this paper, we have reason to believe that the mechanisms of influence on Pinterest are both similar and different to those in other OSNs, and we would like to explore them in greater detail.

\bibliographystyle{aaai}
\bibliography{paper}

\begin{thebibliography}{}

\bibitem[\protect\citeauthoryear{Ahn \bgroup et al\mbox.\egroup
  }{2007}]{ahn2007analysis}
Ahn, Y.-Y.; Han, S.; Kwak, H.; Moon, S.; and Jeong, H.
\newblock 2007.
\newblock Analysis of topological characteristics of huge online social
  networking services.
\newblock In {\em Proceedings of the 16th international conference on World
  Wide Web},  835--844.
\newblock ACM.

\bibitem[\protect\citeauthoryear{Bonneau \bgroup et al\mbox.\egroup
  }{2009}]{bonneau2009eight}
Bonneau, J.; Anderson, J.; Anderson, R.; and Stajano, F.
\newblock 2009.
\newblock Eight friends are enough: social graph approximation via public
  listings.
\newblock In {\em Proceedings of the Second ACM EuroSys Workshop on Social
  Network Systems},  13--18.
\newblock ACM.

\bibitem[\protect\citeauthoryear{Cha \bgroup et al\mbox.\egroup
  }{2010}]{cha2010measuring}
Cha, M.; Haddadi, H.; Benevenuto, F.; and Gummadi, P.~K.
\newblock 2010.
\newblock Measuring user influence in twitter: The million follower fallacy.
\newblock In {\em Proceedings of ICWSM}, volume~10,  10--17.

\bibitem[\protect\citeauthoryear{Chang \bgroup et al\mbox.\egroup
  }{2014}]{chang2014specialization}
Chang, S.; Kumar, V.; Gilbert, E.; and Terveen, L.
\newblock 2014.
\newblock Specialization, homophily, and gender in a social curation site:
  Findings from pinterest.

\bibitem[\protect\citeauthoryear{Chun \bgroup et al\mbox.\egroup
  }{2008}]{chun2008comparison}
Chun, H.; Kwak, H.; Eom, Y.-H.; Ahn, Y.-Y.; Moon, S.; and Jeong, H.
\newblock 2008.
\newblock Comparison of online social relations in volume vs interaction: a
  case study of cyworld.
\newblock In {\em Proceedings of the 8th ACM SIGCOMM conference on Internet
  measurement},  57--70.
\newblock ACM.

\bibitem[\protect\citeauthoryear{Duggan and Smith}{}]{pew}
Duggan, M., and Smith, A.
\newblock Social media update 2013.

\bibitem[\protect\citeauthoryear{Feng \bgroup et al\mbox.\egroup
  }{2013}]{studyuserbehavorspint}
Feng, Z.; Cong, F.; Chen, K.; and Yu, Y.
\newblock 2013.
\newblock An empirical study of user behaviors on pinterest social network.
\newblock In {\em Web Intelligence (WI) and Intelligent Agent Technologies
  (IAT), 2013 IEEE/WIC/ACM International Joint Conferences on}, volume~1,
  402--409.
\newblock IEEE.

\bibitem[\protect\citeauthoryear{Gilbert \bgroup et al\mbox.\egroup
  }{2013}]{statoverviewpint}
Gilbert, E.; Bakhshi, S.; Chang, S.; and Terveen, L.
\newblock 2013.
\newblock I need to try this?: a statistical overview of pinterest.
\newblock In {\em Proceedings of the SIGCHI Conference on Human Factors in
  Computing Systems},  2427--2436.
\newblock ACM.

\bibitem[\protect\citeauthoryear{Gjoka \bgroup et al\mbox.\egroup
  }{2010}]{gjoka2010walking}
Gjoka, M.; Kurant, M.; Butts, C.~T.; and Markopoulou, A.
\newblock 2010.
\newblock Walking in facebook: A case study of unbiased sampling of osns.
\newblock In {\em Proceedings of INFOCOM},  1--9.
\newblock IEEE.

\bibitem[\protect\citeauthoryear{Goodman}{1961}]{goodman1961snowball}
Goodman, L.~A.
\newblock 1961.
\newblock Snowball sampling.
\newblock {\em The Annals of Mathematical Statistics} 32(1):148--170.

\bibitem[\protect\citeauthoryear{Huberman, Romero, and
  Wu}{2008}]{huberman2008social}
Huberman, B.; Romero, D.~M.; and Wu, F.
\newblock 2008.
\newblock Social networks that matter: Twitter under the microscope.
\newblock {\em First Monday} 14(1).

\bibitem[\protect\citeauthoryear{Illenberger, Fl{\"o}tter{\"o}d, and
  Nagel}{2008}]{illenberger2008approach}
Illenberger, J.; Fl{\"o}tter{\"o}d, G.; and Nagel, K.
\newblock 2008.
\newblock An approach to correct biases induced by snowball sampling.
\newblock 26(2010):08--16.

\bibitem[\protect\citeauthoryear{Kamath, Popescu, and Caverlee}{}]{boardrecs}
Kamath, K.~Y.; Popescu, A.-M.; and Caverlee, J.
\newblock Board recommendation in pinterest.

\bibitem[\protect\citeauthoryear{Kurant, Markopoulou, and
  Thiran}{2010}]{kurant2010bias}
Kurant, M.; Markopoulou, A.; and Thiran, P.
\newblock 2010.
\newblock On the bias of bfs (breadth first search).
\newblock In {\em In Proceedings of ITC},  1--8.
\newblock IEEE.

\bibitem[\protect\citeauthoryear{Kurant, Markopoulou, and
  Thiran}{2011}]{kurant2011towards}
Kurant, M.; Markopoulou, A.; and Thiran, P.
\newblock 2011.
\newblock Towards unbiased bfs sampling.
\newblock {\em Selected Areas in Communications, IEEE Journal on}
  29(9):1799--1809.

\bibitem[\protect\citeauthoryear{Lee, Kim, and
  Jeong}{2006}]{lee2006statistical}
Lee, S.~H.; Kim, P.-J.; and Jeong, H.
\newblock 2006.
\newblock Statistical properties of sampled networks.
\newblock {\em Physical Review E} 73(1):016102.

\bibitem[\protect\citeauthoryear{Marlow \bgroup et al\mbox.\egroup
  }{}]{marlow2009maintained}
Marlow, C.; Byron, L.; Lento, T.; and Rosenn, I.
\newblock Maintained relationships on facebook.

\bibitem[\protect\citeauthoryear{Mislove \bgroup et al\mbox.\egroup
  }{2007}]{mislove2007measurement}
Mislove, A.; Marcon, M.; Gummadi, K.~P.; Druschel, P.; and Bhattacharjee, B.
\newblock 2007.
\newblock Measurement and analysis of online social networks.
\newblock In {\em Proceedings of the 7th ACM SIGCOMM conference on Internet
  measurement},  29--42.
\newblock ACM.

\bibitem[\protect\citeauthoryear{Mittal \bgroup et al\mbox.\egroup
  }{2013}]{mittal2013pin}
Mittal, S.; Gupta, N.; Dewan, P.; and Kumaraguru, P.
\newblock 2013.
\newblock The pin-bang theory: Discovering the pinterest world.
\newblock {\em arXiv preprint arXiv:1307.4952}.

\bibitem[\protect\citeauthoryear{neo}{}]{neo4j}
Neo4j the world's leading graph database.

\bibitem[\protect\citeauthoryear{Ottoni \bgroup et al\mbox.\egroup
  }{2013}]{ottoni2013ladies}
Ottoni, R.; Pesce, J.~P.; Las~Casas, D.; Franciscani, G.; Kumaruguru, P.; and
  Almeida, V.
\newblock 2013.
\newblock Ladies first: Analyzing gender roles and behaviors in pinterest.
\newblock {\em Proceedings of ICWSM}.

\bibitem[\protect\citeauthoryear{Romero \bgroup et al\mbox.\egroup
  }{2011}]{romero2011influence}
Romero, D.~M.; Galuba, W.; Asur, S.; and Huberman, B.~A.
\newblock 2011.
\newblock Influence and passivity in social media.
\newblock In {\em Machine learning and knowledge discovery in databases}.
  Springer.
\newblock  18--33.

\bibitem[\protect\citeauthoryear{Suh \bgroup et al\mbox.\egroup
  }{2010}]{suh2010want}
Suh, B.; Hong, L.; Pirolli, P.; and Chi, E.~H.
\newblock 2010.
\newblock Want to be retweeted? large scale analytics on factors impacting
  retweet in twitter network.
\newblock In {\em Proceedings of SocialCom},  177--184.
\newblock IEEE.

\bibitem[\protect\citeauthoryear{Wilson \bgroup et al\mbox.\egroup
  }{2009}]{wilson2009user}
Wilson, C.; Boe, B.; Sala, A.; Puttaswamy, K.~P.; and Zhao, B.~Y.
\newblock 2009.
\newblock User interactions in social networks and their implications.
\newblock In {\em Proceedings of the 4th ACM European conference on Computer
  systems},  205--218.
\newblock Acm.

\bibitem[\protect\citeauthoryear{Wu, Wilkinson, and
  Huberman}{2009}]{wu2009feedback}
Wu, F.; Wilkinson, D.~M.; and Huberman, B.~A.
\newblock 2009.
\newblock Feedback loops of attention in peer production.
\newblock In {\em Computational Science and Engineering, 2009. CSE'09.
  International Conference on}, volume~4,  409--415.
\newblock IEEE.

\bibitem[\protect\citeauthoryear{Zarro, Hall, and
  Forte}{2013}]{zarro2013wedding}
Zarro, M.; Hall, C.; and Forte, A.
\newblock 2013.
\newblock Wedding dresses and wanted criminals: Pinterest. com as an
  infrastructure for repository building.
\newblock In {\em Proceedings of ICWSM}.

\end{thebibliography}

\end{document}